\newcommand{\cc}{\chi^\pm}
\newcommand{\ccp}{\chi^+}
\newcommand{\ccm}{\chi^-}

\newcommand{\cnp}{\chi^\prime}
\newcommand{\sn}{\tilde\nu}

\newcommand{\slr}{\tilde{l}_R}
\newcommand{\sll}{\tilde{l}_L}

\newcommand{\sml}{{\tilde{\mu}_L}}

\newcommand{\str}{{\tilde{\tau}_R}}

\newcommand{\sur}{\tilde{u}_R}
\newcommand{\sul}{\tilde{u}_L}

\newcommand{\sdr}{\tilde{d}_R}
\newcommand{\sdl}{\tilde{d}_L}

\newcommand{\stone}{{\tilde{t}_1}}

\newcommand{\sqs}{\sqrt{s}}

\newcommand{\emis}{{\not{\!\!E}}}
\newcommand{\rpv}{{\not{\!{\mathtt R_p}}}}

\newcommand{\rpc}{{\mathtt R_p}}

\newcommand{\dprim}{{\prime\prime}}
\newcommand{\gevcc}{{GeV/c$^2$}}

\newcommand{\tanb}{{\tan\beta}}

\newcommand{\lbd}{{\lambda}}
\newcommand{\lbdp}{{\lambda^\prime}}
\newcommand{\lbds}{{\lambda^\dprim}}

\documentclass{article}
\usepackage{ltwol2e}

\usepackage{epsfig}
\def\Journal#1#2#3#4{{#1} {\bf #2}, #3 (#4)}

\def\NPB{{\em Nucl. Phys.} B}
\def\PLB{{\em Phys. Lett.}  B}

\def\PRD{{\em Phys. Rev.} D}

\def\PR{{\em Phys. Rep.} }

\def\be{\begin{equation}}
\def\ee{\end{equation}}
\def\bea{\begin{eqnarray}}
\def\eea{\end{eqnarray}}
\begin{document}
\begin{titlepage}

\onecolumn

\begin{picture}(160,1)
\put(350,-50){\parbox[t]{45mm}{{\tt MPI-PhE/98-14}}}
\end{picture}

\vspace{5.cm}

\begin{center}

\rm{\large \bf SEARCH FOR SUSY WITH $\rpv$ AT LEP 
\footnote{
Talk presented at {\it 29$^{th}$ International 
      Conference on High Energy Physics}'', 
      Vancouver, B.C., Canada, July 23-29 1998.
}
}
\vspace{1.cm}

\rm{ G. GANIS }

\vspace{.5cm}
%{\it MPI f\"ur Physik, F\"ohringer Ring 6,
%80805 M\"unchen, GERMANY }
{\it MPI f\"ur Physik, Werner Heisenberg Institut}

{\it F\"ohringer Ring 6, D-80805 M\"unchen, GERMANY }

\end{center}

\vspace{1.cm}

\begin{center}

\begin{tabular}{c}
{\bf Abstract} \\
\multicolumn{1}{p{12.cm}}{
Searches for supersymmetry at LEP allowing for $\rpv$ are reviewed. 
The results are compared with the $\rpc$ conserving scenario.} \\
\end{tabular}
\end{center}

\end{titlepage}

\newpage
% if you want a empty page here ...
\mbox{ }

\newpage

\twocolumn

\title{SEARCH FOR SUSY WITH $\rpv$ AT LEP}  

\author{G. GANIS}

\address{Max Planck Institut f\"ur Physik, F\"ohringer Ring 6,  
80805 M\"unchen, GERMANY\\E-mail: Gerardo.Ganis@cern.ch}   


\twocolumn[\maketitle\abstracts{ 
Searches for supersymmetry at LEP allowing for $\rpv$ are reviewed. 
The results are compared with the $\rpc$ conserving scenario.}]

\section{Introduction}
The searches for supersymmetry performed at LEP  
are guided by a minimal supersymmetric 
extension of the Standard Model~(MSSM)~\cite{HK}, 
based on a minimal number of additional 
degrees of freedom and the addition of soft supersymmetry breaking 
terms, i.e. those breaking the mass degeneracy inside the 
supermultiplets preserving the relevant 
advantages of the theory. Furthermore, in order to insure at tree level 
lepton~($L$) and baryon~($B$) number conservation - which, contrarily to 
the Standard Model, in supersymmetry is not guaranteed by gauge 
invariance - the conservation of the so called $\rpc$ parity 
is required~\cite{rpc}, $\rpc$ being defined as $(-1)^{3B+L+2S}$ with 
$S$ the particle spin.

The so far negative results of these searches have 
pushed the experimental groups to consider less minimal models. 
Among these, those 
obtained allowing some degree of $\rpc$ violation are particularly 
interesting because they predict significant modifications of 
the phenomenology of the reactions searched for.
%, and therefore of the  
%search strategies.

This report summarises the most recent results produced by the LEP 
collaborations on the subject, focusing on the main ideas. 
Details can be found in~\cite{cpapers}.
ALEPH results are based on $\sim$200~pb$^{-1}$ collected at 
$\sqs\!=\!91\!-\!189$~GeV; 
L3 results on $\sim$76~pb$^{-1}$ collected at $\sqs\!=\!161\!-\!183$~GeV;
DELPHI and OPAL results on $\sim$56~pb$^{-1}$ collected at $\sqs\!=\!183$~GeV.

\section{Basic Model and assumptions}

The basic model used as guideline contains the minimal set of 
additional particles: two charginos,~$\cc,\cc_2$, partners of 
$W^\pm$ and $H^\pm$; four neutralinos, 
$\chi,\ \chi^\prime,\ \chi^\dprim,\ \chi^{\dprim\prime}$, partners 
of $\gamma$,~$Z$,~$h,\ H,\ A$; seven complex scalars per family, 
squarks~($\sul,\sdl,\sur,\sdr$), 
sleptons~($\sll^-,\slr^-$), 
sneutrinos~($\sn$), partners of quarks, charged leptons and 
neutrinos.
%% particles as shown in table~\ref{Tab:spectrum}; 
The gluinos~(the eight partners of gluons) and 
the gravitino~(partner of graviton) are assumed to play no r\^ole 
at the energies involved. 
The masses of the Higgs bosons and of the supersymmetric partners
are determined in terms of $\mu$, the Higgs doublet mixing term, 
$\tanb$, the Higgs {\em vev} ratio, $M_1$ and $M_2 = 3/5\cot^2\theta_W M_1$, 
the soft supersymmetry breaking terms associated with $U(1)_Y$ and 
$SU(2)_L$, and $M_0$, a common GUT scale mass for the scalar partners of 
the matter fermions.

The $\rpv$ lagrangian terms can be cast in the form~\cite{rpv}
\begin{equation}
%%{\cal L}_{{\not{{\mathtt R}}}} = 
 \frac{\lambda_{ijk}}{2} L_iL_jE^c_k + \lambda^\prime_{ijk} L_iQ_jD^c_k + 
    \frac{\lambda^\dprim_{ijk}}{2} U^c_iD^c_jD^c_k + \mu_i L_i H_u
\label{Equ:rpv}
\end{equation}
where $i,\ j,\ k$ are generation indices, $L,\ Q$ and $E,\ U,\ D$ 
are, respectively, the lepton and quark $SU(2)$-doublet and 
$SU(2)$-singlet superfields, and $H_u$ is the Higgs doublet 
giving mass to the up-like fermions. Symmetry properties of 
the $\lbd_{ijk}$'s and $\lbds_{ijk}$'s reduce the total number of new 
independent complex parameters to 48. 
The main phenomenological consequence of the $\mu_i$ terms are 
contributions to neutrino masses and oscillations~\cite{hempfling}, 
and are therefore not considered in LEP analyses.

The complexity of the new scenario is reduced by assuming 
that  only one coupling at a time
is non-zero. This is a sufficient 
(although not necessary) condition to cope with the constraints 
derived from low energy experiments on 
the $\lbd$ couplings 
\footnote{Except when the family structure is specified, 
in the following the symbol 
``$\lbd$'' will  denote any of the Yukawa couplings.} 
or on their products~\cite{batta}
\footnote{Typical upper limits on $|\lbd|$'s are of the order 
of a few percent; some products are much more constrained: 
for example, present lower limits on proton lifetime 
require $\lbdp_{11k}\lbds_{11k}<10^{-22}$.}.

In $e^+e^-$ environments the most straightforward consequence 
of the $\rpv$ terms is a modification of the final states deriving 
from pair production of supersymmetric particles.
Non vanishing $\rpv$ couplings allow 
the Lightest Supersymmetric Particle~(LSP) 
to decay into Standard Model~(SM) particles. Consequently, usual cosmological 
arguments demanding a neutral and weakly interacting LSP~\cite{cosmo} do not 
apply any longer. 
The nature of the LSP determines the sensitivity in $\lbd$
 of the searches.
In LEP analyses it is required that the LSP decays within $\sim$1~cm from the 
interaction point. This implies 
$| \lambda |> 10^{-4}$ if the lightest 
neutralino is the LSP, or $| \lambda |> 10^{-7}$ if one of the sfermions  
is the LSP\footnote{Assuming the GUT relation for gaugino masses 
an LSP chargino exists only for $M_{\cc}<M_Z/2$, a mass region ruled out  
by the Z-lineshape measurement; however, it is noted 
that the developed searches cover also chargino LSP topologies.}. 
Lower $|\lbd|$ values give either 
a long-lived LSP, behaving as in the $\rpc$ conserving case, or 
a medium-lived LSP, decaying inside the detector volumes and giving 
topologies not yet addressed
\footnote{Searches for heavy particles decaying inside the detector volume
have been performed by ALEPH and DELPHI~\cite{longlived} 
but not yet interpreted in the 
$\rpv$ scenario.}. 
In the case of a neutralino LSP, for most of the couplings 
the decay length constraint 
can be satisfied only if $M_\chi > 10$~\gevcc.

\begin{figure}
\center
\caption{Typical sensitivity regions in the $(|\lbd|,M_{\tilde{f}})$ plane, 
for current experiments and different experimental techniques. 
Here ``$e^+e^-\ 2f$'' and ``Single Prod'' stand, respectively, 
for two-fermion cross sections and single $\chi$ production in 
$e^+e^-$ interactions.
The gaugino curves come from calculating the $\chi$ 
lifetime assuming that $M_\chi$ is small enough for production at LEP.}
\epsfig{figure=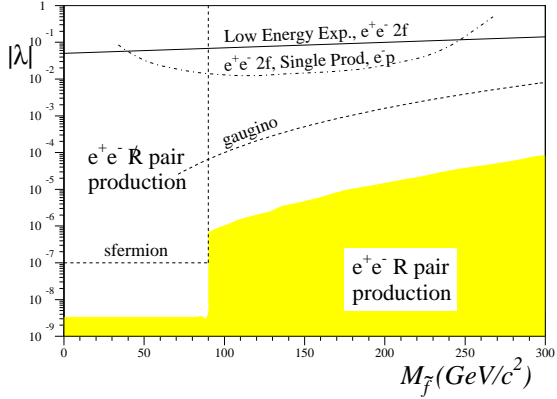,height=6cm}
\label{Fig:lambda_msf}
\end{figure}

The typical experimental situation is summarised in Figure~\ref{Fig:lambda_msf} 
as a function of the relevant parameters: the coupling $\lbd$ and  
the mass $M_{\tilde{f}}$ of the scalar coupling to the $\rpv$ vertex. 
The sensitivity of low energy experiments to large $\lbd$ values 
decreases almost linearly as $M_{\tilde{f}}$ increases. As it will
be discussed briefly in Sect.~\ref{Sec:other}, other signatures 
can be studied at $e^+e^-$ colliders which can be sensitive to lower 
values of $|\lbd|$, for $M_{\tilde{f}}$ below the effective 
centre of mass energy of the experiment.
The same is valid for leptoquark searches at $ep$ colliders~\cite{hera}.  
 Depending on the MSSM 
parameters, pair production in $e^+e^-$ 
can be sensitive to much lower $|\lbd|$ values. 
Also indicated is the region possibly covered by $\rpc$ conserving 
searches.

In the next section the searches for topologies arising 
from pair production are discussed in some detail. 
Single production, example of an alternative signature studied at LEP, 
is discussed in Sect.~\ref{Sec:other}.

\section{Pair Production}

As in the case of $\rpc$ conservation, supersymmetric 
particles are mostly expected to be pair-produced 
via $s$-channel exchange of 
$\gamma,Z$ and, for gaugino and slepton pairs,
 $t$-channel exchange of the relevant slepton or gaugino. 
$\rpv$ couplings involving electrons allow 
additional $t$-channel contributions which, given 
the existing limits on the couplings, are expected 
to be small and neglected. 
Therefore, total production cross sections follow the usual pattern:
typically large for 
gauginos, allowing in some cases to push the sensitivity 
 very close to the kinematic limit; smaller for sfermions,
where the sensitivity is limited by the integrated 
collected luminosity.

\subsection{Decay phenomenology and final state topologies}

The decay modes of the supersymmetric particles mediated by the 
$\rpv$ couplings (the so called {\it direct} decay modes) 
are given in table~1. Depending on the size of 
the couplings and on the nature of the LSP, {\it indirect} decays, 
i.e. those proceeding in a first stage via the lightest neutralino, 
can still play a r\^ole. 
Direct and indirect decays are treated separately since they 
can lead to topologies 
selected with different strategies and performances.

\begin{table}
\label{Tab:decays}
\caption{New direct decay modes of SUSY particles mediated 
by $\rpv$ couplings. Here $i,j,k$ are generations indices.
For lines with more than one particle in the first column, the 
relevant decays are indicated by $\to$ in the second; for example, 
$\tilde{d}_{jL}$ decays via LQD into $\bar{\nu}_i d_k$.}
\begin{center}
\renewcommand{\arraystretch}{1.2}
\begin{tabular}{|c|c|}
\hline
\multicolumn{2}{|c|}{LLE ($\lbd_{ijk}$)} \\
\hline
 $\chi$ & $\to\!\bar{\nu}_i l^+_j l^-_k, \bar{\nu}_j l^+_i l^-_k
           \nu_i l^-_j l^+_k, \nu_j l^-_i l^+_k $    \\  
 $\ccp$ & $\to\!\nu_i\nu_j l^+_k, l^+_i l^+_j l^-_k, 
           l^+_i \nu_j\nu_k, \nu_i l^+_j \nu_k $  \\  
 $\tilde{l}_{iL}^-,\ \tilde{l}_{kR}^-,\ \tilde{\nu}_i$ &  
    $\to\!\nu_j l^-_k,\ \to\!\nu_i l^-_j, \nu_j l^-_i,\ \to\!l^+_j l^-_k $   \\
\hline\hline
\multicolumn{2}{|c|}{LQD ($\lbdp_{ijk}$)} \\
\hline
 $\chi$ & $\to\!l^-_i u_j \bar{d}_k, l^+_i \bar{u}_j d_k, \nu_i d_j \bar{d}_k,  
         \bar{\nu}_i \bar{d}_j d_k$   \\  
 $\ccp$ & $\to\!\nu_i u_j \bar{d}_k, l^+_i\bar{d}_j d_k, l^+_i\bar{u}_j u_k, 
           \bar{\nu}_i \bar{d}_j u_k$ \\  
 $\tilde{d}_{kR},\ \tilde{d}_{jL},\ \tilde{u}_{jL}$ &
  $\to\!\bar{\nu}_i d_j, l^-_i u_j,\ \to\!\bar{\nu}_i d_k,\ \to\!l^+_i d_k$  \\
 $\tilde{l}_{iL}^-,\ \tilde{\nu}_i$ &  $\to\!\bar{u}_j d_k,\ \to\!d_j \bar{d}_k$  \\
\hline\hline
\multicolumn{2}{|c|}{UDD ($\lbds_{ijk}$)} \\
\hline
 $\chi$ & $\to\!u_i d_j d_k, \bar{u}_i \bar{d}_j \bar{d}_k$ \\
 $\ccp$ & $\to\!u_i u_j d_k, u_i d_j u_k, \bar{d}_i \bar{d}_j \bar{d}_k$ \\
 $\tilde{u}_{iR},\ \tilde{d}_{kR},\ \tilde{d}_{jR}$ & 
  $\to\!\bar{d}_j \bar{d}_k,\ \to\!\bar{u}_i \bar{d}_j,\ \to\!\bar{u}_i \bar{d}_k$  \\
\hline
\end{tabular}
\renewcommand{\arraystretch}{1.}
\end{center}
\end{table}

\begin{table*}[t]
\begin{center}
\caption{ Topologies arising from pair production; here 
$l$ stands for any charged lepton, $q$ for any quark and   
$\emis$ indicates the presence of undetected particles. }
\label{Tab:topo}
\renewcommand{\arraystretch}{1.2}
\begin{tabular}{|l|l|l|l|}
\hline
   &   LLE  & LQD & UDD  \\
\hline
 $\ccp\ccm$  & 6$l$, 4$l$+$\emis$, 2$l$+$\emis$, 6$l$+$\emis$,   
            & 2n$q$+2$l$, 2n$q$+$\emis$, 2n$q$+$l$+$\emis${\scriptsize(n=2,3)}, & 6$q$, 8$q$, 10$q$, \\
            &  5$l$+2$q$+$\emis$, 4$l$+4$q$+$\emis$ 
            &  8$q$+2$l$, 8$q$+m$l$+$\emis$ {\scriptsize(m$\le 1$)} & 8$q$+$l$+$\emis$, 6$q$+2$l$+$\emis$ \\
\hline 
 $\chi\chi,\ \chi\cnp$  & 2m$l$+$\emis$ {\scriptsize(m=2,3,4)}, 4$q$+4$l$+$\emis$,   
                & 2n$q$+2m$l$ { ($^{{\mathrm n=2,3,4}}_{{\mathrm m=1,2,3}}$)} 
         & 2n$q$ {\scriptsize(n=3,4,5)}, 2m$q$+$\emis$ {\scriptsize(m=3,4)}, \\
 $\cnp\cnp$ & 2$q$+2m$l$+$\emis$ {\scriptsize(m=2,3)} 
                   & 2n$q$+m$l$+$\emis$ { ($^{{\mathrm n=2,3,4}}_{{\mathrm m\leq 6}}$)}
           & 2n$q$+2m$l$ { ($^{{\mathrm n=3,4}}_{{\mathrm m=1,2}}$)}, 6$q$+2$l$+$\emis$   \\
\hline
 $\tilde{l}^+\tilde{l}^-$  
           & 2$l$+$\emis$, 6$l$+$\emis$ & 4$q$, 4$q$+4$l$, 4$q$+m$l$+$\emis$ {\scriptsize(m=2,3)} & 6$q$+2$l$ \\
 $\sn\sn^*$ & 4$l$, 4$l$+$\emis$ & 4$q$, 4$q$+2$l$+$\emis$, 4$q$+m$l$+$\emis$ {\scriptsize(m=0,1)}  & 6$q$+$\emis$ \\
\hline
 $\tilde{q}\tilde{q}^*$ & 2$q$+4$l$+$\emis$ 
                           & 2$q$+2$l$, 6$q$+2$l$, 2n$q$+m$l$+$\emis$ { ($^{{\mathrm n=1,3}}_{{\mathrm m=0,1}}$)}   
                                                                  & 4$q$, 6$q$, 8$q$ \\
\hline
\end{tabular}
\renewcommand{\arraystretch}{1.}
\end{center}
\end{table*}

%\subsection{Topologies}

Table~\ref{Tab:topo} gives the possible topologies expected 
from pair production. It is noted   
that, despite of the LSP instability, in many cases 
final state neutrinos can carry away 
a significant fraction of energy.

%\subsection{Backgrounds}

\subsection{Selections}

The complexity of the situation depicted in Table~\ref{Tab:topo} 
renders necessary to develop many selections, 
one for every corner of the space of the relevant topological 
and dynamical quantities. 

New signals must be disentangled from the expectations from 
the Standard Model. The most problematic {\it standard} events 
to reject are the 4-fermion events proceeding via heavy gauge bosons 
($W$,$Z$), which, in some cases, result in irreducible contamination 
to be dealt with statistical methods.

% \subsubsection{LLE couplings}

Final states mediated by LLE couplings are characterised by 
a large number of charged leptons and neutrinos. 
The type of events expected 
is quite unusual for the SM; consequently they can be tagged 
efficiently by mainly looking at the number of identified 
leptons.
Hadronic {\it jets} from quark hadronization 
% - hereafter referred to as 
%{\it hadronic jets} or simply as {\it jets} -
can occur in case of indirect decays. In this case the signal events 
are tagged by requiring the identified 
leptons to be isolated from the hadronic system 
and to carry a large fraction of visible energy.

%\subsubsection{LQD couplings}

Direct LQD decays of sleptons or gauginos can lead, respectively, to four jet 
final states  or  four jet and two 
charged leptons final states; in both these cases 
the equal reconstructed mass constraint 
can be exploited to enhance the discriminating power. 
Other decays lead to final states which always 
contain quarks and, depending on the process and the coupling 
structure, charged leptons or neutrinos. Selections therefore 
require hadronic activity, lepton identification and isolation, 
missing energy. Selection efficiencies are generally lower than 
in the LLE case.

% \subsubsection{UDD couplings}

\begin{figure}
\center
\caption{Example of variables used in the selections: reconstructed 
$\chi$ and $\cc$ masses in DELPHI 10-jet analysis.
Data are shown in black dots. The light grey histogram (green in colour) 
corresponds to the expected background. The dark grey (red in colour) 
histogram shows the signal
($\ccp\ccm\!\to\!4j\chi\chi\!\to\!10j$, $M_\cc\!=\!68$~\gevcc and 
$M_\chi\!=\!38$~\gevcc) normalised to 2 pb.}
\vspace{.5cm}
\epsfig{figure=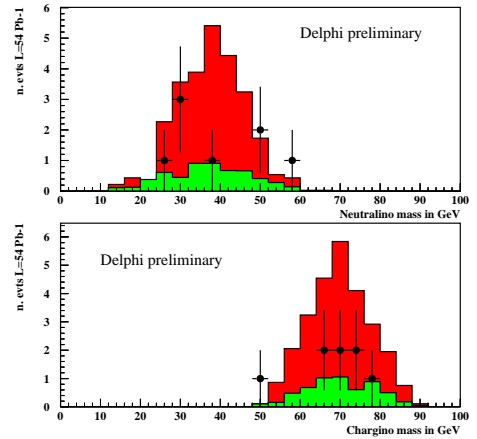,height=5cm}
\label{Fig:mass}
\end{figure}

Direct UDD decays of sleptons or gauginos produce pure hadronic multi-jet 
events, with no missing energy, suffering huge backgrounds 
from two-quark and four-quark SM processes. 
Mass reconstruction and sphericity-like variables are
generally exploited for final discrimination.
Indirect decay modes can lead either to fully hadronic final states 
or to mixed topologies with charged leptons and/or missing energy, 
which can be selected with relatively higher efficiency.
Figure~\ref{Fig:mass} shows an example of variable used 
in these selections.  

Table~3 gives an overlook 
of the typical selections developed by the experimental groups
and of their performances.
%; typical 
%efficiencies are also indicated. The basic features under
%the dominance of the different couplings are now briefly underlined. 

\begin{table}
\label{Tab:sele}
\begin{center}
\caption{Typical selections developed by the experiments~(denoted with their initial)
 in searching for pair production of supersymmetric particles. Here $L$ 
 indicates some degree of charged lepton identifications, 
 $J$ hadronic activity, and $\emis$ missing energy. }
\begin{tabular}{|l|l|l|}
\hline
                    & Exp & Efficiency \\
\hline\hline
 \multicolumn{3}{|c|}{LLE}  \\
\hline
 Acoplanar Leptons                                 & ADLO & $\sim$40\% \\
 ``4$L$'', ``6$L$'' ({\tt w/o} $\emis$)            & ADO  & $\sim$50\%,$\sim$70\% \\
 ``4$L$'', ``6$L$'' + $\emis$                      & ADLO & $\sim$30\%,$\sim$50\% \\
 ``$L+J$'' + $\emis$                                 & ADLO & $\sim$35\%\\
\hline\hline
 \multicolumn{3}{|c|}{LQD} \\
\hline
 ``4J'' ({\tt w/o} $\emis$)                        & ADO & $\sim$50\% \\
 ``4J'' + $\emis$                                  & AD  & $\sim$30\% \\
 ``$L+J$'' ({\tt w/o} $\emis$)                       & ADO & $\sim$30\%\\
 ``$L+J$'' + $\emis$                                 & ADO & $\sim$20\%\\
\hline\hline
 \multicolumn{3}{|c|}{UDD} \\
\hline
 ``6$J$,8$J$,10$J$''                                     & AD & $\sim$15\% \\
 Multi-$J$+$\emis$                                 & A  & $\sim$25\% \\
 ``$L+J$'' ({\tt w/o} $\emis$)                       & A  & $\sim$50\% \\
\hline
\end{tabular}
\end{center}
\end{table}

\subsection{Results}

\begin{figure}
\center
\caption{Event selected by the 4 lepton OPAL LLE search. The most 
likely Standard Model interpretation is
$ZZ\!\to\!e^+e^-\tau^+\tau^-$.}
\epsfig{figure=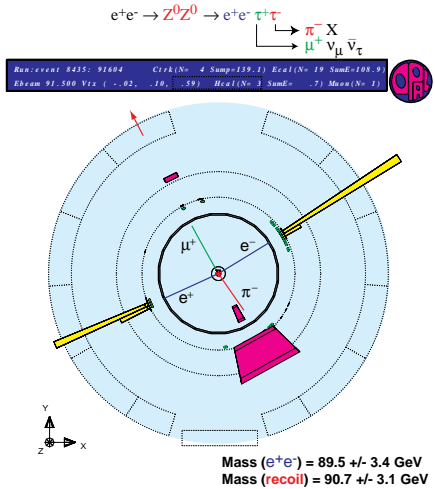,height=6cm}
\label{Fig:cand_4lept}
\end{figure}

As expected by the irreducibility of some of the 
background processes, the  developed analyses select 
candidates. 
%% An example is shown in Figure~\ref{Fig:cand_6jet}. 
An example is shown in Figure~\ref{Fig:cand_4lept}. 
However, the number and dynamics of the selected events is 
in good agreement with the SM expectations. 

The negative results of the searches are used to 
constrain the parameter space. Given the large amount 
of $\rpv$ couplings, a conservative approach is adopted 
to reduce the parameter dependence of the limits by scanning 
the space of the relevant parameters for the most 
conservative configuration;  
for $\lbd$ and $\lbdp$ couplings this corresponds,  in general,   
 to final states with $\tau$'s.

%% \begin{figure}
%% \center
%% \caption{Event selected by the 6 jet ALEPH search. The most 
%% likely Standard Model interpretation is
%% $W^+W^-\!\to\!u_i\bar{d}_j d_k\bar{u}_l + gg$.}
%% \epsfig{figure=aleph_6jet.eps,height=6cm}
%% \label{Fig:cand_6jet}
%% \end{figure}

\subsubsection{Gauginos}

\begin{figure}
\center
\caption{Limits in the $(\mu,M_2)$ plane from ALEPH LLE analysis.}
\epsfig{figure=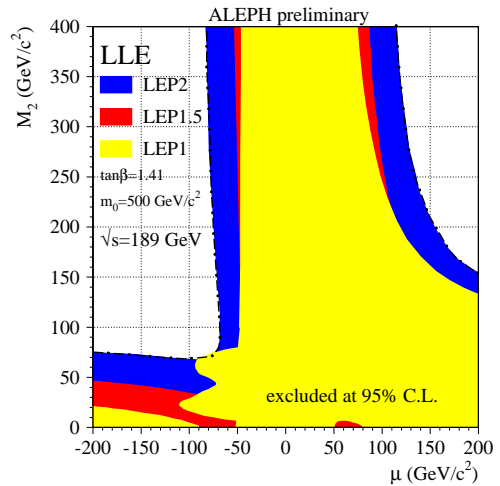,height=6cm}
\label{Fig:mum2}
\end{figure}

For charginos and neutralinos the derived upper limits 
on the production cross section are interpreted in the usual 
$(\mu,M_2)$ plane. The ALEPH result for LLE couplings is 
shown in Figure~\ref{Fig:mum2}, where the first results from 
the 189~GeV data have also been included. 
For large values of $M_0$ the chargino cross section is large 
enough for the excluded region to come very close to 
the kinematic limit, while neutralino cross sections are small and 
can only help when the collected luminosities are large 
\footnote{This is the case, for instance, for LEP~1 results 
that, as can be seen in Figure~\ref{Fig:mum2} begin to be superseded only by 
the most recent data collected at 189~GeV.}. 
Lowering $M_0$, chargino (neutralino) cross sections move 
down (up) because 
of the negative (positive) interference between $t$-channel 
sneutrino (selectron) and $s$-channel exchanges.
The neutralino processes and in particular the visibility 
of $\chi\chi$ act naturally as a backup for the loss 
of sensitivity of $\ccp\ccm$
% rendering the overall 
% picture less dependent on $M_0$
\footnote{ 
The interplay works particularly well 
for LLE, since the topologies arising from $\chi\chi$ 
are exotic and can be selected with high effective 
efficiency. For LQD, the ALEPH analysis with 161 and 172 GeV data 
has shown that the most conservative scenario corresponds to 
dominance of $\chi\chi\!\to\!qqqq\nu\nu$, for which the efficiency 
was not enough to cover, for low $M_0$, the mixed region $\mu\!\sim\!-M_2$; 
assuming similar performances the large luminosity 
collected at 183 is expected to smooth out the $M_0$ dependence. 
The full $M_0$ analysis under UDD dominance has not yet been 
performed, but arguments similar to LQD are expected to hold.}.

\subsubsection{Charged sleptons}

\begin{figure}
\center
\caption{Limits from the charged slepton DELPHI LLE analysis.}
\epsfig{figure=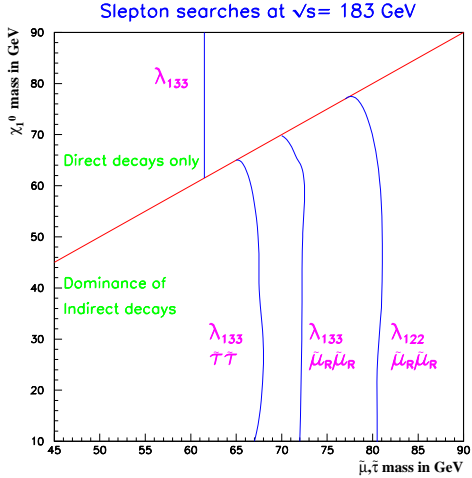,height=6cm}
\label{Fig:slepton}
\end{figure}

In the case of charged sleptons decaying directly into an 
acoplanar lepton pair~(LLE) or four jets~(LQD), the upper 
limit on the cross section is translated into a 
lower limit on the mass. For example, in the 
upper-left part of Figure~\ref{Fig:slepton} it can be seen that 
a $\str^-$ decaying directly in $\tau^-\nu_e$ is excluded by 
DELPHI if its mass is below 62~\gevcc, while from Figure~\ref{Fig:fourjet} 
the ALEPH search for four jet excludes $\sml$ with masses 
below 61~\gevcc. 
Indirect decays, dominating mainly when $\chi$ is the LSP, give 
rise to exotic signatures, like six charged leptons and missing 
energy~(LLE), four quarks and four charged leptons~(LQD), or
six quarks and two leptons~(UDD); these are generally selected 
with better efficiency; the mass lower limits, expressed usually 
in the plane $(M_{\tilde{l}},M_\chi)$, are therefore better than 
in the case of direct decays. The lower part of Figure~\ref{Fig:slepton} 
gives the DELPHI exclusion region for $\tilde{\mu}$ and $\tilde{\tau}$ 
decaying indirectly via $\lbd_{133}$ or $\lbd_{122}$. 
It can be noticed that no unexcluded ``corridor'' is  
present near 
$\Delta M\!=\!M_{\tilde{l}}\!-\!M_\chi\!\sim\!0$ as in the equivalent 
plot for the $\rpc$ conserving case. This is a general feature of 
$\rpv$ searches due to the fact that $\chi$ is visible. However, 
it should be noted that the efficiency is generally lower for small 
$\Delta M$. Therefore 
for cross section limited channels the corridor could still 
appear.

\subsubsection{Sneutrinos}

Direct decays mediated by LLE give the four charged lepton 
signature selected with high efficiency; the worst case are 
$\tilde{\nu}_\mu$'s decaying via $\lbd_{233}$ into $\tau\tau$, 
for which ALEPH gets a lower mass limit of 66~\gevcc. 
LQD direct decays give the four jet signature whose  
non evidence translates in a lower mass limit of 61~\gevcc,
as shown in Figure~\ref{Fig:fourjet}.
Indirect decays lead in this case to less exotic signatures 
and therefore to worse absolute mass limits. For UDD,  
improvements on the Z-lineshape LEP~1 limit are obtained either
for $\tilde{\nu}_e$ in the gaugino region, where constructive 
$t$-channel interference enhances the cross section, or  
considering the three sneutrinos mass degenerate.

\subsubsection{Squarks}

\begin{figure}
\center
\caption{Cross section upper limits from ALEPH LQD 4-jet analysis.}
\epsfig{figure=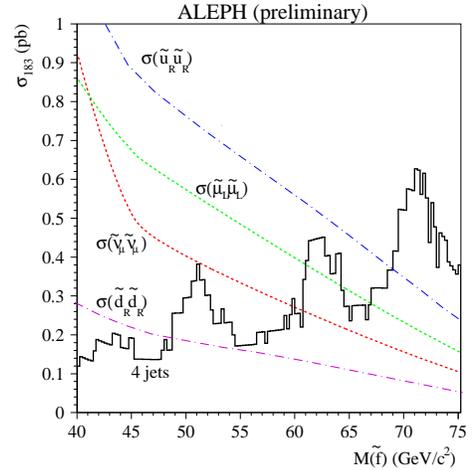,height=6cm}
\label{Fig:fourjet}
\end{figure}

\begin{figure}
\center
\caption{Cross section upper limits from OPAL LQD search 
for $\stone \!\to\!\tau^- d_i$.}
\epsfig{figure=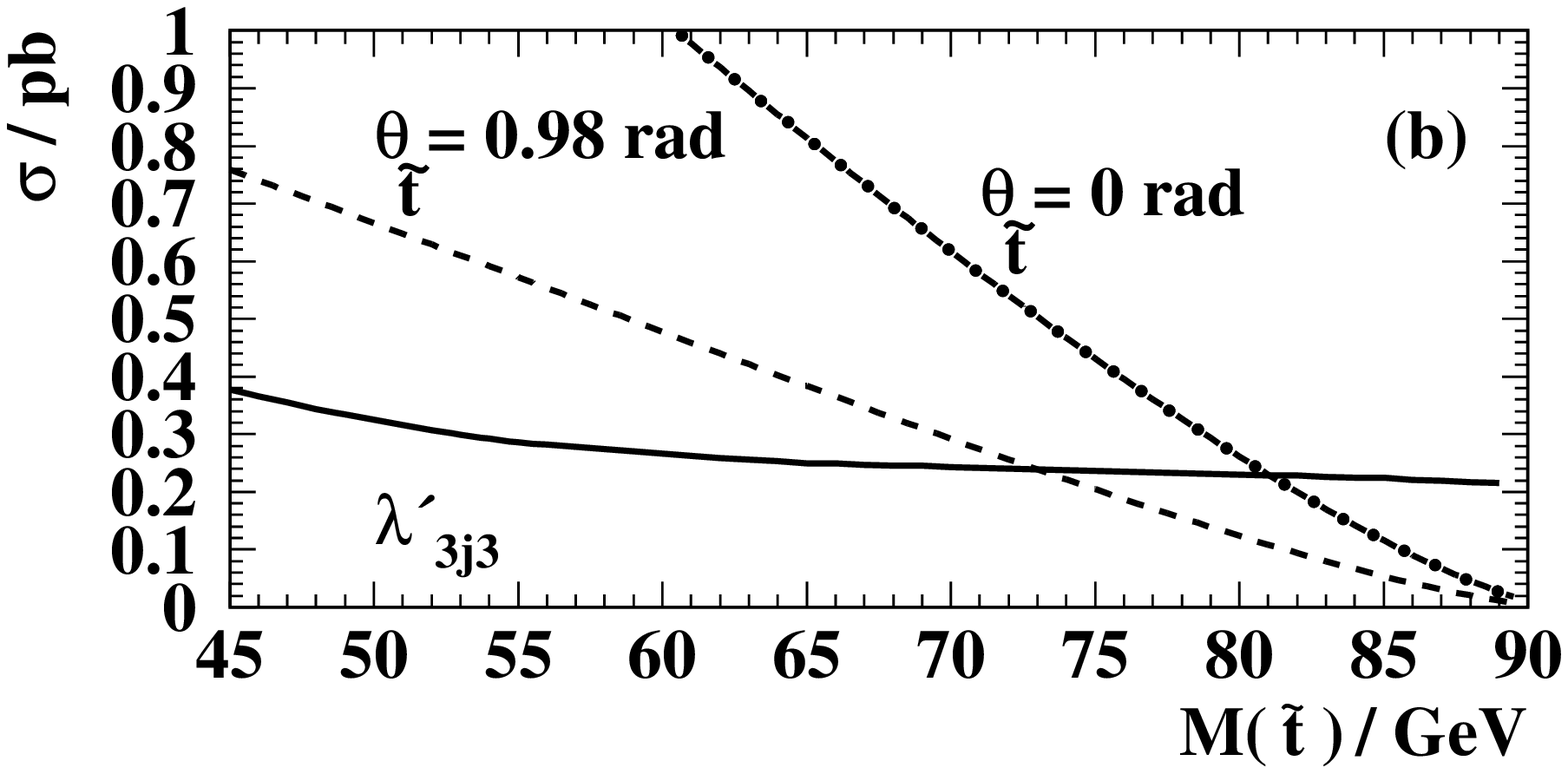,height=4cm}
\label{Fig:stop}
\end{figure}

Direct squark decays via UDD lead to four jet final states. 
As an example, again from Figure~\ref{Fig:fourjet}, ALEPH 
excludes $\sur$'s of masses below 69~\gevcc. LQD direct 
decays give final states with two charged leptons and two 
quarks, which are selected quite efficiently. The OPAL 
result for $\stone \!\to\!\tau^- d_i$, the 
most conservative case, is shown in Figure~\ref{Fig:stop};
for $\stone$ decoupling from the Z the lower mass limit 
is 72~\gevcc.
Topologies arising from indirect decays are selected 
with lower efficiencies, and therefore give worse 
absolute lower mass limits.

It should be noted that hadron colliders are in better 
position for squark searches, especially for LQD 
couplings, where the lepton tag can be used to 
enhance the selection efficiency. First results in this direction 
have been presented at this conference by D0 and CDF~\cite{rpvtev}.

\subsection{Summary of Pair-Production results}

The picture that emerges from 
the study of pair production with $\rpv$ shows that, in 
general, the situation is at least as good as in the 
$\rpc$ conserving case. For gauginos, the $(\mu,M_2)$ analysis for 
gauginos leads to similar conclusions. The typical sfermion limits are  
summarised in Table~4; taking into account 
that the combination of the LEP experiments would bring 
more or less the same gain as in the $\rpc$ conserving case, 
i.e. 5-10 GeV, 
the sensitivity reached allowing for $\rpv$ is very similar 
to the one obtained forbidding it.

\begin{table}
\label{Tab:sfer_sum}
\caption{Summary of lower limits on sfermion masses from pair 
production, in some of the most conservative scenarios. 
Experiments are denoted by their initial. 
DELPHI limits for $\stone$ have been recalculated in the 
most conservative scenario.
Latest limits assuming $\rpc$ 
conservation (labelled as $\rpc$) are also shown for comparison;  
these are taken from the LEP SUSY working group$^9$ for $\stone$ and $\str$,
and from the $Z$ lineshape 
measurement for $\sn$. Values in \gevcc.
}

\begin{center}
\begin{tabular}{|c|c|c|ccc|c|}
\multicolumn{7}{c}{ } \\
\hline
          & Coupling & Decay & A   & D & O & $\rpc$ \\
\hline\hline
 $\sn$    & LLE & indirect & 62 & 62 &    & $\sim$43\\
          & LLE &   direct & 66 & 62 &    & \\
          & LQD &   direct & 60 & 60 &    &   \\
 $\str$   & LLE &   direct & 61 & 61 & 66 &  72 \\
          & LLE & indirect & 70 & 67 &    &     \\
          & LQD &   direct & 62 & 71 &    &   \\
 $\stone$ & LQD &   direct &    & $\sim$71 & 72 &  83.5 \\
          & UDD &   direct & 63 & $\sim$72 &    &   \\
\hline
\end{tabular}
\end{center}
\end{table}

\section{Other Signatures}
\label{Sec:other}

Dominant $\rpv$ couplings whose structure involves the first 
lepton family can manifest themselves in $e^+e^-$ collisions 
in alternative ways.

For example, the $e^+e^-\!\to\!e^+e^-$ cross section is expected 
to receive a contribution from $s$-channel exchange of $\sn_j$ 
if $\lbd_{1j1}\neq 0$; and 
$t$-channel contributions to the hadronic cross section are 
expected from non-zero $\lbdp_{1jk}$ couplings. The study of 
these deviations is part of the electroweak physics at LEP~2 
analyses, and thoroughly discussed in~\cite{ewlep2}.

\begin{figure}
\center
\caption{95\% C.L. cross section upper limits for 
single resonant sneutrino production via $\lbd_{131}$ 
coupling in ALEPH. The rescaling factor $\sigma_{183}/\sigma^\prime$
was calculated explicitly and naturally leads to dips at each 
centre-of-mass energy except 183~GeV.}
\epsfig{figure=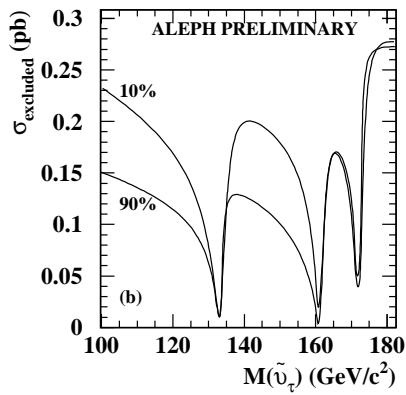,height=5cm}
\label{Fig:singleprod}
\end{figure}

A new analysis has been presented  by ALEPH at this conference 
looking for single neutralino production which would occur 
in the case of non-zero $\lbd_{1j1}$ couplings.
The full reaction would be 
$e^+e^-\!\to\!\tilde{\nu}_j\!\to\!\nu_j\chi\!\to\!\nu_je^+e^-\nu_j$ 
or $\nu_je^+l^-_j\nu_e$, leading to acoplanar $e^+e^-$,  
$e^\pm\mu^\mp$ or $e^\pm\tau^\mp$ final states. 
ALEPH has translated the negative results of 
the acoplanar lepton search into cross section upper limits for 
these processes. The selection efficiency depends 
upon the flavour of the final state leptons, the sneutrino mass 
and the ratio of the neutralino and sneutrino masses. 
In Figure~\ref{Fig:singleprod} the excluded cross section for 
the coupling $\lbd_{131}$ is shown; the best and worst case neutralino 
masses (expressed as a fraction of the sneutrino mass) are plotted. 
Once a point in the MSSM parameter space is chosen, the 
cross section limit can be interpreted as an upper limit 
on $|\lbd_{131}|$; for $\sn$ masses in the range 
of the LEP energies these limits in general 
significantly improve on low energy bounds.

\section{Conclusions}

The LEP collaborations have submitted at this conference 
the results of several searches for supersymmetry allowing 
for some degree of $\rpc$ violation. Overall, these searches 
cover, with good efficiency and purity, most of the signatures 
expected in these scenarios. The resulting 
sensitivity is, in general, as good as 
in the $\rpc$ conserving case, making 
the negative results of searches for supersymmetry at LEP 
not invalidated by $\rpv$.

\section*{Acknowledgements}

I would like to thank the contact persons of the 
LEP collaborations for their prompt help, in
particular I.Fleck and S.Braibant~(OPAL), and 
S.Katsanevas, M.Besancon and C.Berat~(DELPHI). 
I am also indebted with my ALEPH colleagues for 
providing material and 
helpful comments during the preparation of the talk, 
in particular to J.Coles,~P.Coyle,~D.Fouchez and 
M.Williams. Finally I would like to thank 
V.B\"uscher for careful proof-reading of the 
manuscript.

\end{document}